\documentclass[aps,superscriptaddress]{revtex4}
\usepackage{amssymb,amsmath,epsfig}
\usepackage[colorlinks=true, pdfstartview=FitV, linkcolor=blue, citecolor=red, urlcolor=magenta, breaklinks=true]{hyperref}
\usepackage{graphicx}
\usepackage{epstopdf}
\usepackage[capposition=top]{floatrow}
\usepackage{subfig}

\begin{document}
\title{Confinement of fermions in tachyon matter}

\author{Adamu Issifu}\email{bigadamsnet@yahoo.com}
\affiliation{Departamento de F\'isica, Universidade Federal da Para\'iba, 
Caixa Postal 5008, 58051-970 Jo\~ao Pessoa, Para\'iba, Brazil}

\author{Francisco A. Brito}\email{fabrito@df.ufcg.edu.br}
\affiliation{Departamento de F\'isica, Universidade Federal da Para\'iba, 
Caixa Postal 5008, 58051-970 Jo\~ao Pessoa, Para\'iba, Brazil}
\affiliation{Departamento de F\'{\i}sica, Universidade Federal de Campina Grande
Caixa Postal 10071, 58429-900 Campina Grande, Para\'{\i}ba, Brazil}

\begin{abstract}
In this paper we develop a phenomenological model inspired by QCD that mimics QCD theory. We use gauge theory in color dielectric medium ($G(\phi)$) coupled with fermion fields to produce scalar and vector confinement in {\it chromoelectric flux tube} scenario. Abelian theory will be used to approximate the non-Abelian QCD theory in a consistent manner. We will calculate vector and scalar glueballs and compare the result to the existing simulation and experimental results and projections. The QCD-like vacuum associated with the model will be calculated and its behavior studied relative to {changing quark masses}.  
We will also comment on the relationship between tachyon condensation, {dual} Higgs mechanism, QCD monopole condensation and their association with confinement. The behavior of the QCD string tension obtained from the vector potential of the model will be studied to establish vector dominance in confinement theories. 
\end{abstract}

\maketitle
\pretolerance10000

\section{Introduction}
Scalar and vector confinement \cite{2} in $3+1$ dimensional world has been predicted by hadron spectroscopy \cite{1}, confinement in string picture and by QCD lattice simulation but no success has been made in solving it analytically from `first principle' of QCD. It has been shown in quarkonia phenomenology in $3+1$ dimensional world that the best fit for meson spectroscopy are found for a convenient mixture of vector and scalar potentials. 
The combined vector and scalar potential has also been studied in many perspectives. Dirac equation \cite{massive dirac cf} and Schwinger model \cite{schwinger} are among common examples employed  to achieve both linear and coulomb-like potentials \cite{3,4,5,6,7}.  
The attempt to confirm this result predicted by QCD lattice spectroscopy has led to the use of mass gap equation to generate mass as a dependent parameter even in quark systems with no mass current to create dynamic quarks \cite{mas gab}.  
Pure scalar and vector potentials have been dealt with in Refs.\cite{8,9,10,11,12} as a point in focus.

In this work, we use the Lagrangian density for confinement of electric field in tachyon matter \cite{13,14} coupled with fermion fields \cite{14} to produce a Lagrangian density for  fermionic tachyons through transformations. In this approach, the dielectric function coupled to the gauge field and the fermion mass produce the needed strong interaction between (anti-) quark pairs and the scalar field $\phi$ describes the dynamics of the tachyons. We will use a dilaton in gauge theories \cite{15} to determine the coupling constants of the colored particles by transforming the exponential dilaton potential to conform with our chosen tachyon potential considered in this work. 

We show that both scalar and vector confinements coincide with the tachyon condensation. This phenomenon is not completely new because, tachyon fields play a role similar to Higgs fields where Higgs mechanism proceeds via tachyon condensation \cite{17}. Again, both Higgs and tachyon fields have some properties in common, they are both associated with instability or fast decaying with negative mass squared. The tachyons are expected to condense to a value of order of the string scale. Tachyons with large string couplings are considered charged, in that regard, their condensation leads to dual Higgs mechanism. This naturally translates into confinement in line with the QCD monopole condensation scenario \cite{18}. Moreover, we will compute both vector and scalar glueball  masses associated with the model. Glueballs are simply bound states of pure gluons, mixture of quark and gluon(hybrid) and multi quark bound states. The glueball spectrum has been a subject of interest for some decades now, with the focus on unraveling its states in the context of QCD theory \cite{gb1}.    

The success of this type of confinement is based on {\it Nambu-Goto string} or the {\it chromoelectric flux tube picture}. The flux tube scenario is generally observed in static quark frame, such that there is no {\it chromomagnetic field} to induce the spin-orbit interaction of the quarks. The only interaction that is present, in this case, is the {\it kinematics Thomas spin-orbit interaction}, which is a relativistic correction. This is true for both vector and scalar confinement potentials and it is confirmed by lattice simulation results \cite{Thomas interaction}. {\it `Thomas precession'} is considered scalar, hence scalar potential is expected but it does not guarantee its connection with QCD theory. They only relate at long range spin-orbit interactions of QCD, this is precisely the infrared (IR) regime of the theory. The vector potential is achieved for short range Thomas spin-orbit interaction and the scalar potential is consistent with long range Thomas spin-orbit interaction. Both models do not depend on the quarks' spin-orbit interactions in agreement with QCD theory \cite{d vector-scalar}. Though, scalar potential models have been phenomenologically accepted and used in many hadron models, it is still struggling to plant it root firmly in fundamental QCD as highlighted in \cite{Buchmuller,Theodore J. Allen}. Eventually, it was established that in a slowly moving quark frame, QCD predicts both spin dependent \cite{spin-dependent} and spin independent \cite{spin-independent} relativistic corrections.

We will use an Abelian QED throughout the computations, but the color dielectric function $G(r)$ modifies the gauge field and by extension the QCD vacuum of the model \cite{22,23}. We will establish a relationship between the tachyon potential and the color dielectric function in a suitable manner. It is known that the Abelian part of the non-Abelian QCD string tension is 92\%, this represents the linear part of the net potential. Thus, we can make an  approximation of the {non-Abelian} field using an Abelian approach \cite{24,25} --- for recent development on this, see \cite{abelian-1}. Also, it is established that if Abelian projection is followed as suggested by 't Hooft, non-Abelian QCD is reduced to an Abelian theory with charges and monopoles occurring, when such monopoles condense, confinement results. The idea of monopole condensation is a very useful one. It has been numerically shown that monopole condensation actually occurs in the confinement phase of QCD \cite{abelian confinement}. Again, the color dielectric function coupled to the gauge field $F^{\mu\nu}F_{\mu\nu}$, contains only low momentum components. Therefore, the use of Abelian approximation is justified in phenomenological QCD  theory. In sum, these properties enable us to apply phenomenological field theory for QCD in this investigation to establish confinement of quarks and gluons in the infrared regime \cite{4,27,28,29}.

The color dielectric function $G$ is responsible for the long distance dynamics that bring about confinement in the IR regime of the model. It also facilitates the strong interactions between quarks and gluons. The scalar field $\phi(r)$ is responsible for the dynamics of the self-interacting gluon fields and the color dielectric function. 
We will use a Lagrangian density that describes the dynamics of the gauge, the scalar field associated with the tachyon and gluon dynamics and fermion fields coupled with mass at zero temperature \cite{29}. The motivation for using this approach are: Firstly, we are able to compute both the scalar and vector potentials by considering a heavy anti-quark source surrounded by a relatively light and slowly moving quarks. This enables us compute the effect of short range spin dependent {\it `Thomas precession'} (vector potential) and  the effect of long range spin independent  {\it `Thomas precession'} (scalar potential). 
Secondly, we are able to apply phenomenological effective field theory to identify the color dielectric function with the tachyon potential in a simple form. Also, this approach makes it easy to observe how the QCD vacuum is modified by the dielectric function resulting in gluon condensation. Finally, vector and scalar glueball masses are easily calculated from the tachyon potential and the Lagrangian respectively. 

The paper is organized as follows. In Sec.~\ref{sec1} we review the Maxwell's Lagrangian with source in color dielectric medium. In Sec.~\ref{sec2} we introduce  the Lagrangian density for the model. In Sec.~\ref{QCD vacuum} we derive the QCD-like vacuum of the model. In Sec.~\ref{sec3} we choose a suitable tachyon potential for the study. In Sec.~\ref{sec4} we compute the scalar and vector potentials  for confinement, vector and scalar glueball masses and explore the physics involved. In Sec.~\ref{couplings}, we compare our parameters with those from well-known phenomenological models. {In Sec.~\ref{TCDAH} we discuss the dual Higgs mechanism associated with the tachyon (and QCD magnetic monopole) condensation and confinement}.  In Sec.~\ref{sec5} we present the results and analysis. Our final comments are contained in Sec.~\ref{conc}.

\section{Maxwell's Equations Modified by Dielectric function}
\label{sec1}
In this section we will review electromagnetic theory in \textit{color dielectric medium}. Beginning with Maxwell's Lagrangian with source
\begin{equation}\label{1}
\mathcal{L}=-\dfrac{1}{4}F_{\mu\nu} F^{\mu\nu}-j^{\mu}A_{\mu},
\end{equation}
its equations of motion are
\begin{equation}\label{2}
\partial_\mu F^{\mu\nu}=-j^{\nu}.
\end{equation}
The equations of motion of the electromagnetic field create spherical symmetric solutions similar to the well-known point charge scenario associated with Coulomb's field \cite{32}.

Therefore, for electromagnetic field immersed in a dielectric medium $G(\phi)$, where $\phi(r)$ is the scalar field representing the dynamics of the medium, we can construct a Lagrangian density
\begin{equation}\label{3}
\mathcal{L}=-\dfrac{1}{4} G(\phi) F_{\mu\nu} F^{\mu\nu}-j^{\mu}A_{\mu},
\end{equation}
with its equations of motion given as
\begin{equation}\label{4}
\partial_\mu[G(\phi) F^{\mu\nu}]=-j^{\nu}.
\end{equation}
We choose the indices to run as $\mu=1,2,3$ and $\nu=0$. These choices are done carefully to obtain {\it chromoelectric flux} confinement and to eliminate {\it chromomagnetic fields} in the rest frame of the particles. As a result, the equation of motion in Eq.(\ref{4}) can be reduced to
\begin{equation}\label{5}
\nabla.[G(\phi)\textbf{E}]=j^{0}=\rho.
\end{equation}
By this equation, we have eliminated the effect of magnetic field contributions leaving only the electric field contributions that will run through this paper.
From Eq.(\ref{5}), $\textbf{E}$ is the electric field coupled to the dielectric function $ G(\phi)$. Rewriting this equation in spherical coordinates, following the assumption that $E(r)$ and $\phi(r)$ are only functions of $r$, we get
\begin{equation}\label{6}
\nabla.[G(\phi)\textbf{E}]=\dfrac{1}{r^{2}} \dfrac{\partial}{\partial r} (r^{2} G(\phi) E_{r})=\rho.
\end{equation}
Integrating this differential equation yields
\begin{align}\label{8}
&[r^{2}G(\phi)\textbf{E}]=\dfrac{\rho}{\varepsilon_{0}}\int^{R}_{0}r^{2}d r\nonumber\\ &E_{r}=\dfrac{q}{4\pi\varepsilon_{0}r^{2}G(\phi)},
\end{align}
where $q=\dfrac{4}{3}\pi R^{3}\rho$ and $E=\mid\textbf{E}\mid=E_r$. We observe that the dielectric function modifies the magnitude of the electric field function.

\section{Lagrangian Density of the model}
\label{sec2}
In this paper we will focus on a model described by a Lagrangian density given by
\begin{equation}\label{9}
\mathcal{L}=-\dfrac{1}{4}G(\phi)F_{\mu\nu}F^{\mu\nu}+\dfrac{1}{2}\partial_{\mu}\phi\partial^{\mu}\phi-V(\phi)-\bar{\psi}\left(i \gamma^{\mu}\partial_{\mu}+q \gamma^{\mu}A_{\mu}-m_{q\bar{q}}G(\phi)\right)\psi,
\end{equation}
{where $m_{q\bar{q}}$ is the `bare' quark mass or {\it current quark mass} when the particles under consideration are light, and sometimes referred to as the {\it running quark mass} if the particles involved are heavy. It should be seen to be the same as the mass term that appear in the QCD Lagrangian \cite{pdg}. While $M(\phi)=m_{q\bar{q}}G(\phi)$ is the constituent quark mass function which will latter be identified with the scalar potential $S(r)$ in subsequent sections. The `bare' mass ($m_{q\bar{q}}$) leads to {\it explicit chiral symmetry} breaking while the $M(r)$ permits both {\it explicit and dynamical chiral symmetry breaking} similar to renormalized mass in QCD Lagrangian \cite{Melnikov,Heinz}, we will discuss it in details subsequently.}

It should be noted that due to the mass term together with gauge term, $F_{\mu\nu}F^{\mu\nu}$, the Lagrangian is certain to produce both scalar and vector potential contributions to the fermions. The equations of motion of this Lagrangian density are
 \begin{equation}\label{11}
 \partial_{\mu}\partial^{\mu}\phi+\dfrac{1}{4}\dfrac{\partial G(\phi)}{\partial\phi}F^{\mu\nu}F_{\mu\nu}+\dfrac{\partial V(\phi)}{\partial\phi}-\bar{\psi}m_{q\bar{q}}\psi\dfrac{\partial G(\phi)}{\partial\phi}=0,
 \end{equation}

 \begin{equation}\label{13}
 -(i\gamma^{\mu}\partial_{\mu}+q\gamma^{\mu}A_{\mu})\psi+m_{q\bar{q}}G(\phi)\psi=0
 \end{equation}
 
 \begin{equation}\label{15}
 \partial_{\mu}[G(\phi)F^{\mu\nu}]=-\bar{\psi}q\gamma^{\nu}\psi.
 \end{equation}
Again the indices are $\nu=0$ and $\mu=j=1,2,3$ as defined in the previous section. These choices are made deliberately to avoid the creation of {\it chromomagnetic field} so we can focus on the {\it chromoelectric field} which creates the flux tube picture, relevant for our analysis. Therefore, Eq.(\ref{15}) becomes
 \begin{equation}\label{17}
 \nabla.[G(\phi)\textbf{E}] =\bar{\psi}q\gamma^{0}\psi=j^{0}=\rho,
 \end{equation}
where we have substituted $F^{j0}=-E$. Expressing the above equation in spherical coordinates and integrating  the results for electric field solution, yields the same result as Eq.(\ref{8}).


Expanding Eq.(\ref{11}) in radial coordinates to ease our analysis, we get
\begin{equation}\label{20}
-\dfrac{1}{r^{2}}\dfrac{d}{d r}\left[r^{2}\dfrac{d\phi}{d r}\right]-\dfrac{1}{2}\dfrac{\partial G(\phi)}{\partial\phi}E^{2}+\dfrac{\partial V(\phi)}{\partial\phi}-\bar{\psi}m_{q\bar{q}}\psi\dfrac{\partial G(\phi)}{\partial\phi}=0.
\end{equation}
Here we replace $F^{\mu\nu}F_{\mu\nu}=-2E^{2}$ and zero otherwise. This follows from the choice of indices defined above. For simplicity, we will also substitute ($\varepsilon_0=1$) 
\begin{equation}\label{20a}
\lambda=\dfrac{q}{4\pi}
\end{equation}
thus,
\begin{equation}\label{22}
\dfrac{d^{2}\phi}{d r^{2}}+\dfrac{2}{r}\dfrac{d\phi}{d r}=-\dfrac{1}{2}\dfrac{\partial G(\phi)}{\partial\phi}\left[\dfrac{\lambda}{r^{2}G(\phi)}\right] ^{2}+\dfrac{\partial V(\phi)}{\partial\phi}-\bar{\psi}m_{q\bar{q}}\psi\dfrac{\partial G(\phi)}{\partial\phi},
\end{equation}
which implies
\begin{equation}\label{23}
\dfrac{d^{2}\phi}{d r^{2}}+\dfrac{2}{r}\dfrac{d\phi}{d r}=\dfrac{\partial}{\partial\phi}\left[V(\phi)+\dfrac{\lambda^{2}}{2}\dfrac{1}{V(\phi)}\dfrac{1}{r^{4}}-\bar{\psi}m_{q\bar{q}}\psi V(\phi)\right].  
\end{equation}
It has already been established that $G(\phi)=V(\phi)$ --- see \cite{33,Issifu:2017ddb} and references therein --- for slowly varying tachyons. This result will be used throughout this paper.
In the above equation, if we consider a relatively large distance of particle separation from the charge $q$ source, we can ignore the term with $\lambda^2$, hence the equation reduces to
\begin{equation}\label{23c}
\nabla^2\phi=\frac{\partial V(\phi)}{\partial\phi}\left[1-qm_{q\bar{q}}\delta(\vec{r}) \right].
\end{equation}
We have used the general definition $\bar{\psi}\psi\simeq \rho(r)=q\delta(\vec{r})$ in the above equation. 

Switching on the perturbation around the vacuum, $\phi_0=1/\alpha$, i.e. $\phi(r)\rightarrow \phi_0+\eta(r)$, here $\eta(r)$ is a small fluctuation about the true vacuum of the potential, then, Eq.(\ref{23c}) becomes
\begin{align}\label{23d}
&\nabla^2(\phi_0+\eta)=\frac{\partial V(\phi)}{\partial\phi}(1-qm_{q\bar{q}}\delta(\vec{r}))\nonumber\\
&\nabla^2\phi_0+\nabla^2\eta=\left(  \frac{\partial V}{\partial\phi}|_{\phi_0}+\frac{\partial^2 V}{\partial\phi^2}|_{\phi_0}\eta\right) (1-qm_{q\bar{q}}\delta(\vec{r})) \nonumber\\
&\Rightarrow \nabla^2\eta=4\alpha^2(1-qm_{q\bar{q}}\delta(\vec{r}))\eta,
\end{align}
where in the last step we have used Eq.~(\ref{23c}) and anticipated the property of the scalar potential that we shall define shortly.

\subsection{Determining QCD-like Vacuum and Gluon Condensation}\label{QCD vacuum}
This section will be a continuation of the review under `Gluodynamics and QCD-like Vacuum' contained in Ref.\cite{Issifu:2017ddb}. We know that the Lagrangian for gluodynamics is symmetric under conformal transformation, when treated classically i.e. $|\epsilon_v|\rightarrow 0$. Without any quantum corrections, its energy-momentum tensor trace is zero, $\theta^\mu_\mu =0$. As a result, it produces vanishing gluon condensate $\langle F^{\mu\nu}F_{\mu\nu}\rangle= 0$, in the classical limit and non-vanishing gluon condensate $\langle F^{\mu\nu}F_{\mu\nu}\rangle\neq 0$, with quantum corrections. Thus, quantum effects distorts the scale invariance \cite{scale-invariant} and brings about QCD energy-momentum tensor ($\theta^{\mu\nu}$) trace anomaly 
\begin{equation}\label{vg1}
\theta_\mu^\mu=\dfrac{\beta(g)}{2g}F^{a\mu\nu}F^a_{\mu\nu},
\end{equation}
a phenomenon well known in QCD theory. Here, $\beta(g)$ is the QCD beta-function of the strong coupling $g$, with a leading term
\begin{equation}\label{vg2}
\beta(g)=-\dfrac{11 g^3}{(4\pi)^2}.
\end{equation}
This model produces vacuum expectation value
\begin{equation}\label{vg3}
\langle\theta^\mu_\mu\rangle=-4|\epsilon_v|.
\end{equation}
We will now compute the trace of the energy-momentum tensor from the Lagrangian Eq.(\ref{9}) and compare the result with the result obtained above. This comparison is possible because, the third and the forth terms in Eq.(\ref{9}) clearly breaks the scale invariant making it possible for comparison with Eq.(\ref{vg3}). We calculate the energy-momentum tensor trace ($\theta_\mu^\mu$) by substituting Eq.(\ref{11}) into the expression
\begin{equation}\label{vg3a}
\theta_\mu^\mu=4V(\phi)+\phi\square\phi,
\end{equation}
and this yields,
\begin{align}\label{vg3b}
\theta_\mu^\mu &=4V(\phi)-\phi\dfrac{\partial V}{\partial\phi}-\dfrac{\phi}{4}\dfrac{\partial G}{\partial\phi}F^{\mu\nu}F_{\mu\nu}+q\delta({\vec{r})}m_{q\bar{q}}\phi\dfrac{\partial G}{\partial \phi}\nonumber\\
&=4\tilde{V}'+\tilde{G}'F^{\mu\nu}F_{\mu\nu}-4q\delta({\vec{r}})m_{q\bar{q}}\tilde{G}'\nonumber\\
&=4\tilde{V}_{eff}'+\tilde{G}'F_{\mu\nu}F^{\mu\nu}.
\end{align}
Here, we have redefined 
\begin{equation}\label{vg3c}
\tilde{V}'=V-\dfrac{\phi}{4}\dfrac{\partial V}{\partial\phi},\qquad \tilde{G}'=-\dfrac{\phi}{4}\dfrac{\partial G}{\partial\phi}\qquad{\text{and}}\qquad \tilde{V}'_{eff}=\tilde{V}'-q\delta(\vec{r})m_{q\bar{q}}\tilde{G}'.
\end{equation}
Comparing Eq.(\ref{vg3b}) to Eq.(\ref{vg3}) we get,
{
\begin{equation}\label{vg5}
\langle \tilde{G}'(\phi)F^{\mu\nu}F_{\mu\nu}\rangle=-4\langle |\epsilon_v|+\tilde{V}'_{eff}(\phi)\rangle.
\end{equation}
}
We rescale $\tilde{V}'_{eff}(\phi)$ to include the vacuum energy density $-|\epsilon_v|$, i.e.,

\begin{equation}\label{vg6}
\tilde{V}'_{eff}\rightarrow -|\epsilon_v|\tilde{V}'_{eff},
\end{equation}
consequently,
\begin{equation}\label{vg7}
\langle \tilde{G}'(\phi)F^{\mu\nu}F_{\mu\nu}\rangle=4|\epsilon_v|\langle \tilde{V}'_{eff}-1\rangle.
\end{equation}
This equation follows the classical limit, where the gluon condensate vanishes when $|\epsilon_v|\rightarrow 0 $ \cite{gluedynamics}. Again, we will demonstrate in the subsequent sections that, the gluon condensate increases with mass and remains non vanishing at $m_{q\bar{q}}= 0$, i.e. when the quark mass is `removed' after confinement. This is a consequence of {\it chromoelectric flux tube} confinement.

\subsection{Choosing The Appropriate Tachyon Potential}\label{sec3} 
We select a suitable tachyon potential
\begin{equation}\label{23e}
V(\phi)=\frac{1}{2}[(\alpha\phi)^2-1]^2,
\end{equation}
which produces tachyon condensation at low energies. This potential follows the restriction
\begin{equation}\label{potential1}
V(\phi=\phi_0)=0,\qquad \dfrac{\partial V}{\partial\phi}|_{\phi=\phi_0}=0 \qquad\text{and}\qquad \dfrac{\partial V}{\partial \phi}|_{\phi=0}=0.
\end{equation}
These restrictions are necessary to stabilize an asymptotically free system as well as its vacuum \cite{rosina}. To proceed with the computations, we  will require a suitable definition of the three dimensional Dirac delta  function ($\delta(\vec{r})$) that appears in Eq.(\ref{23d}). Consequently, we define $\delta(\vec{r})$ in the limit of step function as \cite{delta function}
\begin{equation}\label{27a}
  \delta(\vec{r})=
\begin{cases}
    \dfrac{1}{4\pi} \lim\limits_{R\to 0} \left(\dfrac{3}{R^{3}}\right), & \text{if}\,\; r\leq R\\
    0,              & r>R
\end{cases},
\end{equation}
where $R$ is the radius of the hadron and $r$ is the inter-particle separation distance. Solving Eq.(\ref{23d}) in the region $r\leq R$, i.e. we are considering the presence of the particles inside the hadron, we get
\begin{equation}\label{27b}
\eta''(r)+\dfrac{2}{r}\eta'(r)+2K\eta(r) =0,
\end{equation}
where $K=2\left(\dfrac{3qm_{q\bar{q}}}{4\pi R^{3}}-1 \right)\alpha^{2}$. This equation has solutions given as
\begin{equation}\label{32}
\eta(r)=\dfrac{\cosh(\sqrt{2|K|}r)}{r\alpha \sqrt{|K|}}\qquad\text{and}\qquad \eta(r)=\dfrac{\sin(\sqrt{2K}r)}{r\alpha\sqrt{K}}, \quad{\text{for}}\quad r\leq R,
\end{equation}
where $|K|=-K=2\left(1-\dfrac{3qm_{q\bar{q}}}{4\pi R^{3}} \right)\alpha^{2}$. {Thus, Eq.(\ref{27b}) has two different solutions each corresponding to different physical regimes. The hyperbolic cosine represents a solution in the low energy regime (IR regime) whilst the sine function presents solution in the high energy regime where the particles are asymptotically free (UV regime) and some contribution from the IR regime where the particles are in a confined state. We will consider these solutions separately as we proceed.} 

Outside the hadron, $r>R$,  $\delta(\vec{r})=0$, we have
\begin{equation}\label{48}
\eta''(r)+\frac{2}{r}\eta'(r)+2K_{0}\eta(r) =0,
\end{equation}
where $K_0=-2\alpha^2$, which has a solution
\begin{equation}\label{50}
\eta(r)=\dfrac{\cosh(\sqrt{2|K_0|}r)}{\alpha r\sqrt{|K_0|}}\qquad\text{and}\qquad \eta(r)=\dfrac{\sin(\sqrt{2 K_0}r)}{\alpha r\sqrt{K_0}}, \quad{\text{for}}\quad r>R.
\end{equation}
{A quark which is kicked out of the pointlike region this way, at relatively high energy, behaves like a massless particle but remains confined to the hadron as would be shown latter}. For now, we will focus our attention on the solution obtained in the region $r\leq R$ (inside the hadron) represented by Eq.(\ref{32}).
The color dielectric function for this solution takes the form
\begin{align}\label{50a}
&G(\phi_0+\eta)=V(\phi_0+\eta)=V(\phi)|_{\phi_0}+V'(\phi)|_{\phi_0}\eta+\frac{1}{2}V''(\phi)|_{\phi_0}\eta^2+{\cal O}(\eta^3)\rightarrow\nonumber\\
&G(\eta)=V(\eta)=\frac{1}{2}V''(\phi)|_{\phi_0}\eta^2.
\end{align}
Tachyon fields are generally unstable with negative mass squared, therefore it is a common phenomenon to find their vacuum states being unstable. In this case, we choose a potential whose true vacuum is at $V(\phi)|_{\phi_0}=0$. Switching on perturbation expansion about $\phi_0$ will results in a generation of square mass proportional to $V''(\phi)|_{\phi_0}$. This stabilizes the tachyon fields and reduce their velocities significantly, making them viable for analysis in the infrared (IR) regime of the model. Simply put, tachyon fields are naturally  associated with instability in quantum field theory hence, we perturb the fields about its true vacuum where the potential has its minima. As a result, our perturbation should be understood as a mechanism to stabilize the tachyon fields \cite{tachyon perturbation}. 

Following from the above equation, the color dielectric function in the IR regime becomes
\begin{align}\label{50b}
G(\eta)&=2\alpha^2\eta^2\nonumber\\
&=\frac{2}{|K|r^2}\cosh^2(\sqrt{2|K|}r).
\end{align}
Substituting the above result for $G(\eta)$ into the electric field equation in Eq.(\ref{8}) we get
\begin{align}\label{35}
E&=\dfrac{\lambda}{r^{2}G}\nonumber\\
&=\dfrac{\lambda}{r^{2}\left[\dfrac{2}{r^2|K|}\cosh^2(\sqrt{2|K|}r)\right]}.
\end{align}
The same tachyon potential and the procedure adopted here was used in \cite{Issifu:2017ddb}  to confine light quarks at a finite temperature. Consequently, the tachyons are expected to generate mass at $V''(\phi)|_{\phi_0}$ leading to a particle-like state resulting in tachyon condensation \cite{30}.  
It follows that, the vacuum stability is independent of the constituent quark mass appearing in the Lagrangian density. These properties observed from this type of potential makes it more efficient for computing confinement potentials and glueball masses for heavy or light (anti-)quark systems. From the afore analysis, we can associate $f_\alpha=1/\alpha$ to the {\it decay constant} of the tachyons, the higher its value the faster the tachyons decay and by extension, the more unstable the QCD-like vacuum created in this process and the vice versa.

\subsection{Potentials, Glueball Masses and Constituent Quark Masses}\label{sec4}
\subsubsection{The potential of the particles inside the hadron}\label{sec4a}
Using the well-known relation for calculating electromagnetic potentials 
\begin{equation}\label{v-p}
V_{c}(r)=\mp\int Edr, 
\end{equation}
to determine the confinement potential $V_{c}(r,m_{q\bar{q}})$. {Now we apply the hyperbolic cosine function at the left side of Eq.(\ref{32}) representing the solution in the IR regime
}
to get
\begin{align}\label{35a}
V_c(r,m_{q\bar{q}})&=\mp\frac{\lambda\sqrt{|K|}\tanh[\sqrt{2|K|}r]}{2\sqrt{2}}+c\nonumber\\
&=\mp\frac{\lambda\sqrt{(1-\beta qm_{q\bar{q}})}\alpha\tanh[\sqrt{4(1-\beta qm_{q\bar{q}})}\alpha r]}{2}+c,
\end{align}
here, $\beta$ is a constant representing the depth of the delta function well, i.e. $\beta=\delta(R\rightarrow 0)={3}/{(4\pi R^{3})}$. The above equation gives the net potential observed by massive (anti-)particle pairs ((anti-)quark pairs). 
In effect, particles of the same kind (two particles or two anti particles) repel each other whilst particles of different kinds (particle and anti-particle) attract; for example, color (blue) attracts an anti-color (anti-blue) of the same kind or colors of diffident kinds (blue and green) attracts each other and the vice versa \cite{8}. 

We choose the negative part of the potential which correspond to the potential of an anti-particle. It is appropriate to choose  $q=-1$ also corresponding to an anti-charge. We will latter find that $q=-1$ is associated with an anti-charge which is identifiable with an anti-color charge carried by the gluons. Considering that, we are dealing with the potential of an anti-particle, this choice is justifiable. 
Also, we will choose $R=(3/4\pi)^{1/3}$ i.e. $\beta=1$ and $c=0$ ($c$ is the integration constant). Consequently, the  net static potential (a combined scalar and vector potentials) seen by an antiparticle confinement in the region $r\leq R$, in the rest frame of a heavy source is given by
\begin{equation}\label{38}
V_c(r,m)=-\frac{\lambda\sqrt{(1+m_{q\bar{q}})}\alpha\tanh[\sqrt{4(1+m_{q\bar{q}})}\alpha r]}{2}, 
\end{equation}
and its string tension is given as$^1$\footnotetext[1]{The parameters combined as follows are in general related according to the form $\Big(m_{q\bar{q}}+1/(2\pi\alpha')^{1/2}\Big)\tilde{\alpha}^2$, where $\tilde{\alpha}^2=\alpha^2/(2\pi\alpha')^{1/2}$. This is because the precise relationship between tachyon potential and color dielectric function is $G(\phi)=(2\pi\alpha')^2V(\phi)$, where $\alpha'$ is the Regge slope which has dimension of length squared. However, we have assumed $2\pi\alpha'=1$ along the paper.} 
\begin{equation}\label{39}
\sigma_c(m_{q\bar{q}})=-\lambda(m_{q\bar{q}}+1)\alpha^2=(m_{q\bar{q}}+1)\alpha^2.
\end{equation}
{The scalar potential contribution here is due to the heavy point-like source situated at the center of the hadron.} { In this regime or at relatively large inter particle separations $r$, particles are always confined with hadron degrees of freedom. This result does not present UV effects, i.e., Coulombic potential effect at $r\rightarrow 0$. If we choose $\sigma_c\sim 1\text{GeV/fm}$ and $\alpha=0.99$ such that $f_\alpha\simeq 1$ we obtain $m_{q\bar{q}}=10.10\text{MeV}$, which lies within the range of u and d-quark masses. It is important to add that this potential is good for investigating light quarks such as u, d and s-quarks, since the potential already lies in the stable regime of the theory where all the quarks are confined into hadrons. Heavy quarks such as c-quarks starts experiencing the degeneracy (pair production) as shown in Fig.~\ref{fig:4} with dashed lines. Therefore, this model is recommended for confining light quarks such as u, d or s only. 
}
 
 {This result accounts for both spin-dependent and spin-independent relativistic corrections as QCD predicts for slowly moving quark systems. Because there are spinless quarks (comparatively light) in a slow motion around the heavy point-like anti-quark source at the origin. The co-motion of the light quarks relative to the central massive static anti-quark source creates spin-dependent corrections at short ranges (vector-type interaction) and spin-independent correction at long distances (scalar-type interaction) \cite{quark motion} giving rise to the net potential.} 
 To this end, when the surrounding co-moving quarks are far away from the heavy source, the string tension that binds them to the massive anti-quark source will break leading to pair production (a (anti-)quark pair creation), a phenomenon well known in QCD theory. However, in the presence of a heavy anti-quark source, the new quark created  in the process remain attractive towards the original source whereas the new anti-quark created in the process remain attractive towards the `break-away' quark. In this regard, the (anti-)quark pairs will always remain confined once there exist, at least, a heavy source. The process remains the same if we consider a heavy quark source as well. This characteristic is depicted by the linear nature of the string tension changing with $m_{q\bar{q}}$ and constant when $m_{q\bar{q}}=0$.

{We will now analyse the behaviour of the particles in the UV regime where the particles are relatively close i.e. small inter particle separation $r$, with quark and gluon degrees of freedom. 
We will use the solution of the sine function at the right side of Eq.(\ref{32}) for this analysis, 
  \begin{align}\label{uv1}
\eta(r)&=\dfrac{\sin(\sqrt{2K}r)}{\alpha r\sqrt{K}}.
 \end{align}
Substituting this result into the color dielectric function Eq.(\ref{50b}) results in,
 \begin{equation}\label{uv2}
 G_s(r)=\dfrac{2}{K r^2}\sin^2(\sqrt{2K}r).
 \end{equation}
Substituting the above equation into  Eq.(\ref{v-p}),
 \begin{equation}
V_s=\int\dfrac{\lambda}{2r^2\left(\dfrac{\sin(\sqrt{2K}r)}{r\sqrt{K}}\right)^2}dr
\end{equation}
we get
\begin{align}\label{cornell}
V_s(m_{q\bar{q}},r)&=-\dfrac{\lambda\sqrt{K}\cot[\sqrt{2K}r]}{2\sqrt{2}}+\tilde{c} \nonumber\\
&\simeq -\dfrac{1}{4r}+\dfrac{K}{6}r+\tilde{c}\nonumber\\
&\simeq -\dfrac{1}{4r}+\dfrac{(m_{q\bar{q}}-1)\alpha^2}{3}r+{\cal O}(r^3) +\tilde{c},
\end{align}
with string tension 
\begin{equation}\label{cornella}
\sigma_s=\dfrac{(m_{q\bar{q}}-1)\alpha^2}{3}.
\end{equation}
Where we have chosen the positive part of the potential corresponding to $\lambda=q=1 $ and $\tilde{c} =0$. 
It is known in perturbation QCD that the dominant interaction at small distances, $r\rightarrow 0$, is Coulombic one-gluon exchange (OGE). 

We can estimate the {\it running quark mass}, ($m_{q\bar{q}}$), using this result, if we consider a typical hadron of mass $1\text{GeV}$ with radius $1\text{fm}$ as determined in electron scattering with estimated string tension $\sigma_{s}\sim 1\text{GeV/fm}$ and a decay of $f_\alpha=1$. We can estimate the {\it running quark mass} as $m_{q\bar{q}}=4\text{GeV}$, a typical mass for heavy quarks. 
The mass is slightly less than the mass of a b-quark, $m_b=4.18\pm 0.03\text{GeV}$ on mass-independent subtraction scheme ($\overline{MS}$) at a scale of $\mu=2\text{GeV}$ as reported in the Review of Particle Physics of the Particle Data Group \cite{pdg}. In addition, light quark masses (such as u, d and s-quark masses) are difficult to determine due to their small masses relative to hadron scale, so  it is sometimes difficult to classify and significantly identify their influence within hadrons. Thus, this model is vaible for investigating confinement of quarks with masses $m_{q\bar{q}}\geq 4\text{GeV}$, bellow this threshold the quarks are expected to be asyptotically free in the UV regime. It is therefore, convenient to study the behaviour of light quarks using the IR potential developed in Eq.(\ref{38}), instead of the Cornell-like potential in Eq.(\ref{cornell}) known for confining heavy quarks \cite{conell:1}.
\subsubsection{String Tension}
Generally, the  Cornell potential for confining heavy quarks is giving by 
\begin{equation}\label{con1}
V(r)=-\dfrac{e}{r}+\sigma r+V_0,
\end{equation}
where $e$ and $\sigma$ are the Cornell fit parameters, but $\sigma$ is related to the lattice spacing $a$ as $\sigma\sim 1/a^2$ and $V_0$ is the self energy of the static source. The separation distance, $r$, between the particles determine the magnitude of  $a$, i.e. large distances imply wide $a$ and vice versa \cite{conell:1}. Determination of $e$ is strictly phenomenological, its larger value correspond to smaller mass and the vice versa. 
In relation to our model, we normalized it at $V_0=\tilde{c}=0$, this is to remove the divergence known to be caused by self-energy contribution in the continuum limit and $e=0.25$ \cite{Gunnar}. Unlike the quenched approximation where the particle and antiparticle numbers are independently conserved, in this model the conserved quantity is the baryon number due to the presence of the {\it current quark mass} and a {\it constituent quark mass function}. As would be shown explicitly later, light quark-antiquark pairs are created in the vacuum resulting from the transition in the string tension connecting the two static sources leading to hadronization. When the energy carried by the string tension connecting the two static sources exceed its critical value at some separation distance $r=r_*$ or some critical mass $m_{q\bar{q}}=m_c$, the string will `break' and decay into light static mesons. Thus, the Coulomb effect $e$ is rather pronounced for higher masses and weaker if the mass involved is slightly weaker than the infrared mass --- see the results published in Ref. \cite{Aoki}. Thus in the limit $r\rightarrow r_*$ the potentials stop rising with distance $r$ and the static source quark becomes screened by the light quarks formed in the vacuum. Similar behaviour is observed when we keep the distance fixed and increase the mass i.e. $m_{q\bar{q}}\rightarrow m_c$. 
At the ground state $\sigma_c>\sigma_s$. To obtain confinement, the $\sigma_c$ can contain current mass within the range $0\leq m_{q\bar{q}}\leq 10.10\text{MeV}$ whereas $\sigma_s$ can contain current mass within the range $4\leq m_{q\bar{q}}\leq m_c\text{GeV}$ due to the UV contributions. The running masses for the IR and the UV regimes can be compared with the results of the running masses calculated using various QCD sum rules within the size of a hadron. The masses are giving as $\bar{m}_u(\text{1GeV})=5.2\pm 0.5\text{MeV}$, $\bar{m}_d(\text{1GeV})=9.2\pm 0.5\text{MeV}$, $\bar{m}_s(\text{1GeV})=159.5\pm 8.8\text{MeV}$ and $\bar{m}_b(1\text{GeV})=5.8\pm 0.06\text{GeV}$ \cite{Narison,Hatsuda}.
   }

\subsubsection{Vector Potential}\label{sec.v}

{Using the solution at the left side of Eq.(\ref{50}) i.e. outside the regime of the pointlike source, we will have two separate potential contributions; {\it vector potential} due to the gluonic sector and the massless quark and a scalar potential energy contribution from the system of the pointlike particles (hadron). Here, the hadron serves as a massive point-like source that confines the massless quark.  We will compute the {\it vector potential} in this section while we reserve the calculations of the {\it scalar potential energy} and the {\it net potential energy} for the next section. Using Eqs.(\ref{35}) and (\ref{v-p}), the vector potential becomes,

\begin{align}\label{a}
V_v(r)&=\mp\dfrac{\lambda\sqrt{|K_0|} \tanh[\sqrt{2|K_0|}r]}{2\sqrt{2}}+c\nonumber\\
&=\mp\dfrac{\lambda\alpha\tanh(2r\alpha)}{2}+c,
\end{align}
with string tension
\begin{equation}\label{a1}
\sigma_v=\mp\lambda\alpha^2=\mp\dfrac{\lambda}{f^2_\alpha}.
\end{equation}
This corresponds to the short range Thomas spin-orbit interactions as highlighted in the introduction. Again, faster tachyon decay means weak string tension while slow decay means strong string tension and stronger confinement. Similar results is obtained if we set $m_{q\bar{q}}=0$ in Eq.(\ref{35a}).

}

\subsubsection{Scalar Potential}\label{scp}

To determine the resulting scalar potential due to the hadron, we compare our results from Eq.(\ref{13}) with the generalized Dirac equation where the scalar and the vector potentials  coexist as
\begin{equation}\label{2.79a}
[ c\hat{\alpha} \hat{p} + \hat{\beta} m_{0} c^{2} +V(r)] \psi =0 .
\end{equation}
Here, we have used $P^\mu=(\frac{E}{c},\vec{P})$, $A^\mu= (\Phi(r), \vec{A})$, assuming that the quarks are scalar (spinless) and static, we have  $E=\vec{A}=\Phi(r)=0$. In spherical wall potential, we can impose the restriction
 \begin{equation}\label{2.791}
 V(r)= \begin{cases}
    S(r) ,& \text{for } r\leq R,\\
    0, & \text{for } r>R \end{cases}
     \end{equation}
where $r$ is the inter-quark separations and $R$ is the radius of the hadron. Here, $V(r)=S(r)$ represents the scalar potential in the Dirac equation. Thus, Eq.(~\ref{2.79a}) can be rewritten as
 \begin{equation}\label{2.79b}
( c\hat{\alpha} \hat{p} + \hat{\beta} m_{0} c^{2})\psi +S(r) \psi =0,
\end{equation}
where $\hat{\alpha}$ and $\hat{\beta}$ are 
Dirac matrices.
Also, $\hat{p}\rightarrow -i\nabla$ ($\hbar=1$) is the momentum operator, $m_0$ is the rest mass of the fermions and $c$ is the speed of light in vacuum. The vector potential is normally introduced by minimal substitution of the momentum $P_\mu\rightarrow P_\mu-gA_\mu$ whilst the scalar potential is introduced by the mass term $m\rightarrow m_0+S$, $g$ is a real coupling constant. It should be noted that the vector and scalar potentials are coupled differently in the Dirac equation \cite{dirac equation}.
Comparing Eq.(\ref{13}) and Eq.(\ref{2.79b}), we can identify the first and second terms of both equations as the interaction terms and the third terms as the scalar potentials \cite{ada:2,ada:3}. We leave out the vector potential because it has already been calculated in Sec.~\ref{sec.v} so, it is of no further interest. The resulting scalar potential seen by the fermions is
\begin{align}\label{40}
S(r,m_{q\bar{q}})= m_{q\bar{q}}G(r)=\frac{1}{2}m_{q\bar{q}}V''(\phi)|_{\phi_0}\eta^2(r)=2m_{q\bar{q}}\alpha^2\eta^2=m_{q\bar{q}}\left[\frac{2}{r^2|K|}\cosh^2(\sqrt{2|K|}r) \right].
\end{align}
{This result represents Thomas spin-orbit interactions at long ranges, where the short range Thomas spin-orbit interactions are partially dominated giving way to scalar interactions only. As a result, the interactions are thought of as being concentrated on the various quark coordinates}. 
Interestingly, the {\it scalar potential energy} is simply a product of $m_{q\bar{q}}$ and the color dielectric function. This gives an indication that confinement in this scenario has a direct relation with tachyon condensation and $m_{q\bar{q}}$ of the system. 

Meanwhile, some authors have predicted vector dominance over scalar for inter quark potentials, suggesting that scalar dominance will imply that all quark combinations are confined. Contrary to that, only the energetic combinations are preferred phenomenologically \cite{8}. We will attempt to analyse this assertion by computing and comparing the magnitudes of their coupling constants as they are expected to appear in the Dirac equation and its significance. 
{The net confining potential energy of a quark outside the hadron will be the sum of  Eqs.(\ref{a}) and (\ref{40}) yielding 
\begin{equation}\label{c}
V_{\text{net}}(r,m)=\pm\dfrac{q^2\alpha\tanh(2\alpha r)}{8\pi}+\dfrac{m_{q\bar{q}}}{r^2(1+m_{q\bar{q}})\alpha^2}\cosh^2(\sqrt{4(1+m_{q\bar{q}})}\alpha r).
\end{equation} 
Here, we have multiply the vector potential in Eq.(\ref{a}) with an anti-particle charge $-q$ to obtain a vector potential energy that goes into the net potential energy.
 } 

{To end this section, we will present the color dielectric function and the scalar potential for the sine function used in investigating the particles inside the hadron in the UV regime. Consequently, the color dielectric function will be 
\begin{align}\label{colorc}
G_s(r,m_{q\bar{q}})&= \left[ 4-\dfrac{8K}{3}r^2\right] \nonumber\\
&=  \left[ 4-\dfrac{16(m_{q\bar{q}}-1)\alpha^2}{3}r^2\right], 
\end{align}
and the scalar potential of the hadron reads 
\begin{align}\label{scalarc}
S_s(r,m_{q\bar{q}})&=m_{q\bar{q}}\left[ 4-\dfrac{8K}{3}r^2\right] \nonumber\\
&= m_{q\bar{q}}\left[ 4-\dfrac{16(m_{q\bar{q}}-1)\alpha^2}{3}r^2\right], 
\end{align}
for $\lambda=q=1$.

}

\subsubsection{Glueball Masses}
This model is certain to produce both vector and scalar glueballs just as vector and scalar potentials calculated above. Scalar glueball mass has been estimated to have a value within the range $1.5$ to $1.7\text{GeV}$. This value has been affirmed by data and calculations \cite{scalar glueball}. Specific value was reached in Ref.~\cite{scalar glueball 1} to be $1.7 \text{GeV}$ for fitness, through unquenched calculation. 
 On the other hand, the existence of vector glueball mass has been predicted with an extimated value of $3.8\text{GeV}$ by quenched Lattice QCD \cite{vector glueball}. The vector glueball masses are expected to be observed at the ongoing Beijing Spectrometer Experiment (BESIII) and hopefully future PANDA experiment at the FAIR Lab \cite{vector glueball 1}. Recent findings published by BESIII facility \cite{experimental report} points to vector particles of mass $3.77$ to $4.60\text{GeV}$ with precision. 

From the tachyon potential in Eq.(\ref{23e}) we can directly calculate the `vector glueball'  mass ($m_{gb}$) as
\begin{equation}\label{gb}
m_{gb}^2=\dfrac{\partial^2 V}{\partial\phi^2}|_{\phi_0}=4\alpha^2.
\end{equation}
The scalar glueball mass is then calculated directly from the Lagrangian in Eq.(\ref{9}) as
\begin{align}\label{gb1}
m_{gb\phi}^2&=-\left\langle \dfrac{\partial^2\mathcal{L}}{\partial\phi^2}\right\rangle |_{\phi_0}=4\alpha^2(1+m_{q\bar{q}})+\alpha^2\langle F^{\mu\nu}F_{\mu\nu}\rangle|_{\phi_0}\nonumber\\
&=4\sigma_c,
\end{align}
where, we have substituted $\psi\bar{\psi} =q\delta(\vec{r})\rightarrow q\beta$, $q=-1$ and $\beta=1$. Following the analysis in Sec.~\ref{QCD vacuum}, we find that $\langle F^{\mu\nu}F_{\mu\nu}\rangle|_{\phi_0}=0$. Therefore, the model produces vanishing gluon condensate in its vacuum ($\phi_0$). 
We can now rewrite Eq.(\ref{gb1}) in terms of the string tension $\sigma_c$ and the vector glueball mass as,
{
\begin{equation}\label{gb2}
m_{gb\phi}^2=\dfrac{m_{gb}^2\sigma_c}{\alpha^2}=m_{gb}^2\sigma_cf^2_\alpha.
\end{equation} 
}
{From the afore analyses, we find that the vector and scalar glueballs depend on the tachyon decay constant, $\alpha=1/f_\alpha$. Whilst the vector glueball depends directly on $f_\alpha$ the scalar glueball only depends on $f_\alpha$ through the string tension $\sigma_c$}. 
Knowing the estimated values of scalar and the vector glueball masses and the value of the string tension $\sigma_c\sim 1\text{GeV/fm}$ 
we can fix the tachyon decay constant within the range, $0<f_\alpha\leq 1$. If we consider that the tachyons are decaying at half of its maximum value, i.e. $f_\alpha=0.5$, we will obtain $m_{gb\phi}=2\text{GeV}$ and $m_{gb}=4\text{GeV}$. {This decay regime gives a good agreement {with} the vector glueball mass within the range of the experimental values. That notwithstanding, vector and scalar glueballs are more likely to be observed for slowly decaying tachyons.} 

\subsubsection{Constituent quark masses}
{We would like to make brief comments on the constituent quark mass function $M(\phi)$ that was mentioned bellow Eq.(\ref{9}). This function is the same as the scalar potential energy $S(r)$ determined from the Dirac's equation and discussed under Sec.~\ref{scp}. Here, we will seek to determine its value for $r\rightarrow 0$ and $r\rightarrow r_*$ i.e. the excited state and the ground state respectively. 

Now expanding the solution for the hyperbolic cosine in Eq.(\ref{32}) for $\sqrt{ 2|K|}r\ll 1 $, we get
\begin{align}\label{constm1}
\eta(r)&=\dfrac{\cosh(\sqrt{2|K|}r)}{r\alpha \sqrt{|K|}}\nonumber\\
&\simeq \dfrac{1}{r\alpha\sqrt{|K|}}\left[ 1+|K|r^{2}\right]+{\cal O}(r^{3}) .
\end{align}
Using the expression for $G(\eta)$ in Eq.(\ref{50b}) and the above solution, we can write 
\begin{align}\label{constm2}
M(r)&=\dfrac{2m_{q\bar{q}}}{r^{2}|K|}\left[1+2|K|r^{2} \right] +{\cal O}(r^{3})\nonumber\\
&=\dfrac{2m_{q\bar{q}}}{r^{2}|K|}+4m_{q\bar{q}} +{\cal O}(r^{3}).
\end{align}
Thus, the constituent quark mass for the ground state energy will be $M(r\rightarrow r_*)=4m_{q\bar{q}}$ \cite{Gunnar}, where $r_*$ represents the distance within which the quarks are confined and beyond {it} we have degeneracy. Therefore, the constituent quark mass ($M_{r_*}$) for the potential in Eq.(\ref{38}) is $40.4\text{MeV}$, and degeneracy should be expected beyond this mass limit. 

We can now proceed to equivalently calculate the constituent quark mass ($M_0$) for the highest excited state $r\rightarrow 0$ by using the sine function solution in Eq.(\ref{32}), this yields
 \begin{equation}\label{constm21}
M(r)=4m_{q\bar{q}}\left[1-\dfrac{K r^{2}}{3} \right] +{\cal O}(r^{3}).
\end{equation}
Therefore, $M(r\rightarrow 0)=4m_{q\bar{q}}$ is the constituent quark mass in this regime, if we choose a running quark mass of $m_{q\bar{q}}=4\text{GeV}$ as calculated above, we get, $M_0=16\text{GeV}$. So, the constituent  mass ($M$) in this model framework lies between $0\leq M\leq 16 \text{GeV}$, where $0\leq M_{r_*}\leq 40.4\text{MeV}$ gives the dynamics of the IR regime while the UV regime can be studied within $4\leq M_0\leq 16\text{GeV}$.

The quark masses that appear in phenomenological models are generally the {\it constituent quark mass} \cite{pdg} which dynamically breaks the chiral symmetry. In non perturbative theories the range for \textit{dynamical chiral symmetry} breaking, $\Lambda_\chi$ is about $1\text{GeV}$ \cite{chiral-a}. It is therefore conventional to say a quark is heavy if $M>\Lambda_\chi$, in this case, we have {\it explicit chiral symmetry breaking} for c, b and t-quarks whilst quarks are classified as light if  $M<\Lambda_\chi$, leading to {\it spontaneous chiral symmetry breaking} dominance. Quarks in this category include u, d and s. In nonrelativistic quark models, one of the useful parameters is the constituent quark mass which is determined to be $M_u=M_d=350\text{MeV}$ for light quarks in {\it single gluon exchange} (OGE) interaction, while the predicted mass for a c-quark is also $M_c=1.6\text{GeV}$. Constituent quark mass models are used to study the effect of {\it dynamical chiral symmetry breaking} 
and they are independent of the {\it current quark mass}. The {\it constituent quark mass} changes depending on how measurements are made in a particular model framework. The constituent quark masses are  greater compared to {\it current quark mass} and are usually free parameters to fit in potential models. The masses vary depending on whether you are using a meson fits, baryon fits, hadron fits or other phenomenological models.
}  

\subsection{Identification of Coupling Constants}
\label{couplings}
In this section, we will compare the result from equation (\ref{23}) with the well-known phenomenological models \cite{8,34}, using dilaton in gauge theory to confine quarks and gluons with $N_{c}$ colors. For emphasis, since our model is phenomenological, it lacks all the degrees of freedom required to fully represent pure QCD theory, but our results agree with the QCD spectrum for scalar and vector potentials.

The results presented above still keeps the electromagnetic charge $q$ and do not contain the color numbers $N_c$. This comparison is intended to help fill in these gaps. Furthermore, the scalar and the vector potentials in Dirac equation are of the same weight but coupled differently. This comparison will enable us to determine their couplings and relate them to their individual strengths in the QCD-like model. The same method was used in \cite{33} --- and references herein --- to determine couplings as well. The dilaton model is given as
\begin{equation}\label{41}
\dfrac{d^{2}\phi}{d r^{2}}+\dfrac{2}{r}\dfrac{d\phi}{d r}=-\dfrac{g^{2}}{64\pi^{2}f_{\phi}}\left( 1-\dfrac{1}{N_{c}}\right) \exp\left(- \dfrac{\phi}{f_{\phi}}\right)\dfrac{1}{r^{4}}-\dfrac{\xi}{2f_{\phi}}\exp\left( -\xi\dfrac{\phi}{2f_{\phi}}\right)m_{q\bar{q}} g \delta(\textbf{r}). 
\end{equation}
Transforming the exponential potentials of the above equation to conform with the tachyon potential used, i.e,
$\exp\left(-\dfrac{\phi(r)}{f_{\phi}}\right)\rightarrow2(\alpha^{4}\phi^{3}-\alpha^{2}\phi)$ and $\exp\left(-\xi\dfrac{\phi(r)}{2f_{\phi}}\right)\rightarrow2\left(\dfrac{\alpha^{4}\phi^{3}\xi^{4}}{16}-\dfrac{\alpha^{2}\phi\xi^{2}}{4} \right)$,
hence equation (\ref{41}) can be rewritten as
\begin{equation}\label{42}
\dfrac{d^{2}\phi}{d r^{2}}+\dfrac{2}{r}\dfrac{d\phi}{d r}=\dfrac{g^{2}}{32\pi^{2}f_{\phi}}\left(1-\dfrac{1}{N_{c}} \right)\dfrac{\alpha^{2}\phi}{r^{4}}+\dfrac{2\xi^{3}\alpha^{2}\phi}{8f_{\phi}}m_{q\bar{q}} g \delta(\textbf{r}). 
\end{equation}
It should be noted that the purpose of this section is to determine the coupling constants, so the equations have been simplified to achieve that goal.

Comparing equations (\ref{23}) and (\ref{42}), we can find
$\xi=2$,  and $\alpha^2={1}/{f_{\phi}}={1}/{f^2_\alpha}$. Here $\xi$ represents the coupling strength of the fermions, $f_\phi$ is the mass decay constant of the dilaton and it is related to the decay constant of the tachyons $f_\alpha$, $g$ is the gluon charge and it is related to the electromagnetic charge $q$. We can rewrite equations (\ref{20a}), (\ref{38}) and (\ref{39}) as
\begin{align}\label{44a}
&\lambda=-\dfrac{g}{2\pi}\left(1-\dfrac{1}{N_c} \right)^{1/2}\implies\nonumber\\ &
q=-g \sqrt{\left( 1-\dfrac{1}{N_c}\right)},
\end{align}
\begin{align}\label{44}
 V_{c}(r,m_{q\bar{q}})&=\dfrac{g}{4\pi}\sqrt{\left(1-\dfrac{1}{N_{c}} \right)}\frac{\sqrt{(1+m_{q\bar{q}})}\alpha\tanh[\sqrt{4(1+m_{q\bar{q}})}\alpha r]}{2},
 \end{align}
 and
\begin{equation}\label{45}
\sigma_c(m_{q\bar{q}})=\dfrac{g}{4\pi}\sqrt{\left(1-\dfrac{1}{N_{c}} \right)}(m_{q\bar{q}}+1)\alpha
\end{equation}
for $g\rightarrow g/(2\sqrt{2})$ respectively.

The {\it scalar potential energy} observed by the heavy pointlike anti-quark source is also given as
\begin{equation}\label{scalar p}
S(r,m_{q\bar{q}})=2m_{q\bar{q}}\alpha^2\eta(r)^2=m_{q\bar{q}}\left[\frac{1}{r^2(1+m_{q\bar{q}})\alpha}\cosh^2(2\sqrt{(1+m_{q\bar{q}})}\alpha r) \right].
\end{equation}
We find from the above equation (\ref{50b}) for the dielectric function $G$ that, the larger the quark mass the more stable the QCD-like vacuum and the faster the tachyons condense \cite{35}. Since tachyon condensation implies confinement, then heavier quarks are more likely to be confined than light quarks \cite{ada:4}.
This potential follow the same analysis as the dielectric function with the tachyons condensing faster in a multiple of $m_{q\bar{q}}$ showing a rather stronger confinement but vanishes at $m_{q\bar{q}}=0$. This result represents the energy flow between a quark and an anti-quark pairs at long ranges. On the other hand, Eq.(\ref{44}) represents the {\textit{net potential}} for confinement observed for a heavy anti-quark source in the origin surrounded by relatively light spinless quarks in a slow motion.

It is important to state that, the net confinement potential seen in Eq.(\ref{44}) corresponds to a mixture of $\xi=0$ (no mass coupling) and $\xi=1$ (Kaluza-Klein type or mass coupling) \cite{coupling:1}. The combined $\xi=1$ and $\xi=0$ gives the net confinement of the quarks at all masses inside the hadron. Also, Eq.(\ref{scalar p}) gives the \textit{scalar potential energy}, i.e., the flow of energy between two or more quarks. The coupling of the scalar potential, $\xi=2$, exposes its weakness relative to the vector potential in Eq.(\ref{a}) by a ratio of $0:2$. { By simple interpretation, the scalar potential energy must be coupled strongly in order to coexist with the vector potential energy which needs no coupling as seen in our model framework and presented in expression Eq.(\ref{c}).} It is noticeable that the scalar potential energy vanishes at $m_{q\bar{q}}=0$ while the vector potential energy does not as evident in Eq.(\ref{c}). 
{Consequently, quarks are always in a confined state \cite{chiral-s:1} under the net confining potential Eq.(\ref{44}) and at $m_{q\bar{q}}=0$ we retrieve the vector potential Eq.(\ref{a}).}  
This indicates that the {\it chromoelectric flux} generated by the colored particles remain confined even if we 'remove' the quarks after confinement under the net confining potential ($V_c$) \cite{1}. It should be noted also that, the net confinement potential of the color particles is the sum of the vector and the scalar potentials as presented in (\ref{44}) and the net potential energy is also the sum of the vector and scalar potential energies as expressed in  Eq.(\ref{c}). 

{After knowing all the coupling constants and the nature of the Dirac delta function, few comments on Sec.~\ref{QCD vacuum} will be necessary. Noting that the Dirac delta function is well defined inside the hadron $r\leqslant R$, with depth $\beta=3/(4\pi R^3)=1$, $G(\phi)=V(\phi)$ and an anti-gluon color charge $q\rightarrow g=-1$  $(N_c\gg1)$, Eq.(\ref{vg3c}) takes the form
\begin{align}\label{vg5b1}
\tilde{V}'_{eff}(\phi)&=\tilde{V}'-q\delta(r)m_{\bar{q}q}\tilde{G}'\nonumber\\
&=V-\dfrac{\phi}{4}\dfrac{\partial V}{\partial \phi}+m_{q\bar{q}}\left( -\dfrac{\phi}{4}\dfrac{\partial G}{\partial \phi} \right) \nonumber\\
&=V-\dfrac{\phi}{4}\dfrac{\partial V}{\partial \phi}(m_{q\bar{q}}+1)
\end{align}
Substituting Eq.(\ref{50a}), recalling that  $V(\eta(r))=G(\eta(r))=2\alpha^2\eta^2$ for $\phi(r)\rightarrow\phi_0+\eta(r)$, into the above expression yields
\begin{align}\label{vgb2}
\tilde{V}'_{eff}&=2\alpha^2\eta^2-(\phi_0+\eta)\alpha^2\eta(m_{q\bar{q}}+1)\nonumber\\
&=\alpha^2\eta^2(1-m_{q\bar{q}})-\phi_0\alpha^2\eta(1+m_{q\bar{q}})\nonumber\\
&=\alpha^2\eta^2((1-m_{q\bar{q}})-\alpha\eta(1+m_{q\bar{q}})\qquad{\text{for}}\qquad \alpha=1/\phi_0.
\end{align}
Since tachyon condensation is faster when $\tilde{V}'_{eff}\rightarrow 0$, see Fig.~\ref{fig:6}, tachyons condense with increasing mass and attain its minimum condensation at $m_{q\bar{q}}=0$ corresponding to the maximum value of $\tilde{V}'$.
As a results, Eq.(\ref{vg7}) yields,

\begin{align}\label{vg7a}
&\langle (\phi_0+\eta){G}'(\eta)F_{\mu\nu}F^{\mu\nu}\rangle=16|\epsilon_v|\langle 1-\alpha^2\eta^2(1-m_{q\bar{q}}) \rangle+16|\epsilon_v|\langle\alpha\eta(1+m_{q\bar{q}})\rangle \rightarrow\nonumber\\
&\langle (\alpha\eta+\alpha^2\eta^2)F_{\mu\nu}F^{\mu\nu}\rangle=4|\epsilon_v|\langle 1-\alpha^2\eta^2(1-m_{q\bar{q}}) \rangle+4|\epsilon_v|\langle\alpha\eta(1+m_{q\bar{q}})\rangle.
\end{align}
By this result, tachyon condensation increases with increasing mass $m_{q\bar{q}}$, and attain its minimum when $m_{q\bar{q}}=0$. There is a QCD monopole condensation associated with both instances (i.e. $m_{q\bar{q}}=0$ and $m_{q\bar{q}}>0$ ), since the gluon condensate do not vanish in any of the two instances. The condensate is higher with increasing mass and relatively low at no mass. Naturally, confinement results in both instances. 
 
\subsection{Tachyon Condensation and Dual Abelian Higgs Mechanism}
\label{TCDAH}

For the sake of simplicity, but without lost of generality, in the previous sections we have postponed the discussion concerning the dual description of the confinement in terms of the Higgs mechanism. We shall now complete the discussion by extending the the Lagrangian in Eq.(\ref{9}) by imposing gauge invariance on the scalar sector. As such, the degrees of freedom should be carefully changed. Firstly we have to consider a charged scalar field, i.e., a complex scalar field
\begin{eqnarray}
\phi=\frac{\phi_1+i\phi_2 }{\sqrt{2}}.
\end{eqnarray}

We will leave out the fermion (spinors) coupling from the original Lagrangian in Eq.(\ref{9}) for this analyses because it has no direct influence on the outcome of the intended result. Consequently,
\begin{equation}\label{hm2}
\mathcal{L}=-\dfrac{1}{4}G(|\phi|){F}_{\mu\nu}{F}^{\mu\nu}+D_{\mu}\phi D^{\mu}\phi^\ast-\dfrac{1}{4}\tilde{F}_{\mu\nu}\tilde{F}^{\mu\nu}-V(|\phi|),
\end{equation}
where $F_{\mu\nu}=\partial_\mu A_\nu-\partial_\nu A_\mu$ and $\tilde{F}_{\mu\nu}=\partial_\mu \tilde{A}_\nu-\partial_\nu \tilde{A}_\mu$ are two independent Abelian field strengths and $D_\mu =\partial_\mu-iq\tilde{A}_\mu$ is the Abelian covariant derivative associated with the {\it dual} gauge field $\tilde{A}_\mu$ which is responsible for the magnetic monopole description in the dual Higgs mechanism. 
Besides its original $U(1)$ gauge invariance, the Lagrangian becomes invariant under the following $\tilde{U}(1)$ gauge transformation
\begin{align}\label{hm2b}
&\phi(x)\rightarrow\phi'(x)=e^{iq\alpha(x)}\phi\nonumber\\
&\tilde{A}(x)\rightarrow \tilde{A}'(x)=\tilde{A}(x)-\partial_\mu\alpha(x).
\end{align}

We can now analyze the dual Higgs mechanism of the model. 
Going forward, the potential of the model in Eq.({\ref{23e}) will also take the form
\begin{equation}\label{hm3}
V(|\phi|)=2\alpha^4(\phi\phi^\ast)^2-2\alpha^2\phi\phi^\ast+\dfrac{1}{2}.
\end{equation}
The minima of this potential function is given as 
\begin{align}\label{hm4}
&\dfrac{\partial V}{\partial\phi^*}=0\rightarrow 4\alpha^4(\phi\phi^\ast)\phi-2\alpha^2\phi= 0,\nonumber\\
&\langle |\phi|\rangle_0= \sqrt{\phi_0\phi_0^\ast}=\sqrt{\dfrac{1}{2\alpha^2}}.
\end{align}
The non-zero vacuum expectation value breaks the $\tilde{U}(1)$ gauge symmetry spontaneously.   

Let us now proceed as in our previous analyses by expanding about the vacuum of the potential so, for ease of calculations, we choose physical vacuum configuration
\begin{equation}\label{hm5}
\phi_{1min}=\phi_0 \qquad\text{and}\qquad \phi_{2min}=0.
\end{equation} 
Now, we choose two real scalar fields $\eta(r)$ and $\zeta(r)$  \cite{chris quigg,griffiths} to represent small fluctuations about the vacuum of the potential. 

So we can represent the shifted vacuum as
\begin{equation}\label{hm8}
\phi=\eta+\langle\phi\rangle_0.
\end{equation}  
 As such, we can conveniently parametrize the $\phi$ field as 
 \begin{align}\label{hm9}
 \phi&=e^{i\zeta/\phi_0}\dfrac{(\phi_0+\eta)}{\sqrt{2}}\nonumber\\
 &\approx \dfrac{(\phi_0+\eta(r)+i\zeta(r))}{\sqrt{2}}.
 \end{align}
The Lagrangian for the small fluctuations about the vacuum up to quadratic order becomes 
\begin{equation}\label{hm10}
\mathcal{L}=\dfrac{1}{2}[\partial_\mu\eta\partial^\mu\eta-4\alpha^2\eta^2]+\dfrac{1}{2}[\partial_\mu\zeta\partial^\mu\zeta-2q\phi_0 \tilde{A}_\mu\partial^\mu\zeta+{q^2\phi^2_0}\tilde{A}_\mu \tilde{A}^\mu]-\dfrac{1}{4}G(\eta)F_{\mu\nu}F^{\mu\nu}-\dfrac{1}{4}\tilde{F}_{\mu\nu}\tilde{F}^{\mu\nu}+...
\end{equation}
Notice that the scalar field $\eta$ plays the role of the usual massive Abelian Higgs field and the scalar field $\zeta$ is related to the massless Goldstone boson. Now choosing the gauge
 \begin{equation}\label{hm12}
 \tilde{A}_\mu\rightarrow \tilde{A}'=\tilde{A}_\mu-\dfrac{1}{q\phi_0}\partial_\mu\zeta,
 \end{equation}
we can write
\begin{align}\label{hm13}
\dfrac{1}{2}\partial_\mu\zeta\partial^\mu\zeta-q\phi_0 \tilde{A}_\mu\partial^\mu\zeta+\dfrac{q^2\phi^2_0}{2}\tilde{A}_\mu \tilde{A}^\mu&=\dfrac{q^2\phi_0^2}{2}\left[ \tilde{A}_\mu-\dfrac{1}{q\phi_0}\partial_\mu\zeta\right] \left[ \tilde{A}^\mu-\dfrac{1}{q\phi_0}\partial^\mu\zeta\right] \nonumber\\
&=\dfrac{q^2\phi_0^2}{2}\tilde{A}'_\mu \tilde{A}'^\mu.
\end{align}
Thus, the 
 Lagrangian takes the form 
\begin{equation}\label{hm14}
\mathcal{L}=-\dfrac{1}{4}G(\eta)F_{\mu\nu}F^{\mu\nu}+\dfrac{1}{2}\partial_\mu\eta\partial^\mu\eta-V(\eta)-\dfrac{1}{4}\tilde{F}'_{\mu\nu}\tilde{F}'^{\mu\nu}+\dfrac{q^2\phi^2_0}{2}\tilde{A}'_\mu \tilde{A}'^\mu+...
\end{equation}
This is precisely the Lagrangian (\ref{9}) evaluated around the vacuum plus the Lagrangian of the dual gauge field. Notice that $G(\eta)=V(\eta)=2\alpha^2\eta^2$ is consistent with (\ref{50a}).
Furthermore, as we have well discussed in the previous sections, in the limit $\eta\to0$ we have tachyon condensation and confinement. Then at such a limit, we are left with the Lagrangian
\begin{equation}\label{hm14-v2}
\tilde{\mathcal{L}}=-\dfrac{1}{4}\tilde{F}'_{\mu\nu}\tilde{F}'^{\mu\nu}+\dfrac{q^2\phi^2_0}{2}\tilde{A}'_\mu \tilde{A}'^\mu,
\end{equation}
which describes the dynamics of a massive dual gauge field --- a consequence of the dual Higgs mechanism.

The equations of motion of the dual gauge field are 
\begin{equation}\label{hmc}
\partial_\mu \tilde{F}'^{\mu\nu}=q^2\phi_0^2\tilde{A}'^\nu,
\end{equation} 
and that for the static fields give
\begin{equation}\label{hmc-v2}
\nabla\cdot\vec{\tilde{B}}=\rho_m \quad\text{and}\quad -\nabla\times\vec{\tilde{E}}=\vec{j}_m,
\end{equation} 
where the magnetic monopole charge and current densities $\rho_m$, $\vec{j}_m$, respectively, are defined by the current density $j^\mu=q^2\phi_0^2\tilde{A}'^\mu$. 
So these equations govern a dual superconductor with the dual London equation $\nabla\times\vec{j}_m=(1/\lambda^2)\vec{\tilde{E}}$,  where  $\lambda=(q^2\phi_0^2)^{-1/2}$ is the penetration depth \cite{monopole condensation 1}.  This scenario leads to the description of condensed magnetic monopoles and confined electric field, as we previously assumed.

}}

\section{Results and Analysis}
\label{sec5}
\begin{figure}[H]
  \centering
  \subfloat[Left Panel]{\includegraphics[scale=0.5]{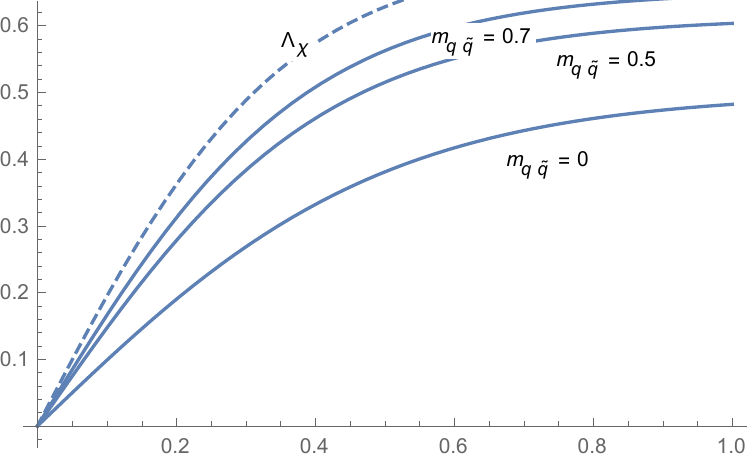}}
  \qquad
  \subfloat[Right Panel]{\includegraphics[scale=0.5]{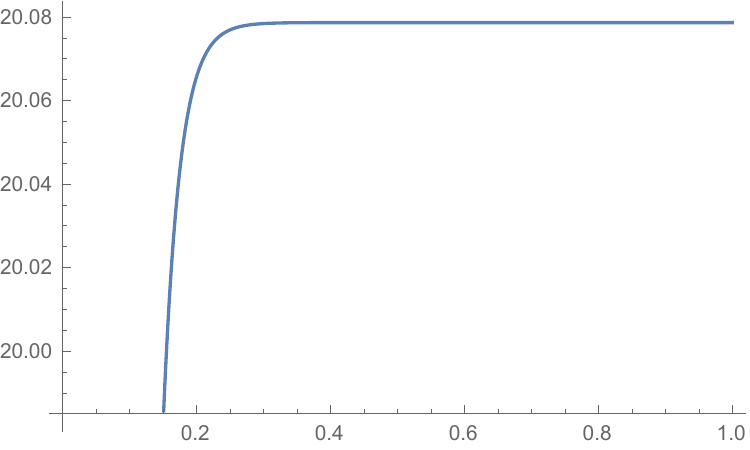}}
  
   \caption{A graph of net potential, $V_{c}(r,m_{q\bar{q}})$, against ($r,m_{q\bar{q}}$) for specific values of $m_{q\bar{q}}$ (left) and an infinite $m_{q\bar{q}}$ (right).}
   \label{fig:4}
   \floatfoot{The graph {in the left panel} is for different values of $m_{q\bar{q}}$. It shows the strength of $q\bar{q}$ confinement for different values of $m_{q\bar{q}}$ as their separation distance $r$ vary. The graph shows a steady increase in its gradient as $m_{q\bar{q}}$ is increased from $0$ to $0.7\text{GeV}$. By inference, the particles are strongly confined with increasing mass. That notwithstanding, a system of fermions remain confined even if the mass of the fermions is  `removed' ($m_{q\bar{q}}=0$) after confinement due to the \textit{chromoelectric flux tube confinement}, in consistence with the QCD theory. {The graph in the right panel is for infinite limit of quark masses, it rises sharply, indicating confinement and flatten up shortly indicating hadronization. $\Lambda_\chi = 1\text{GeV}$ is the threshold mass bellow which quarks are classified as light and beyond which they are classified as heavy.}}
\end{figure}

\begin{figure}[H]
  \centerline{\includegraphics[scale=0.5]{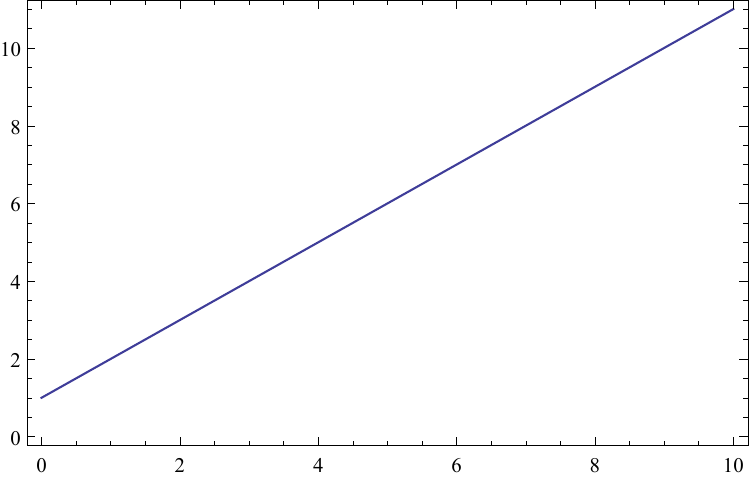}}
  \caption{A graph of string tension $\sigma_c(m_{q\bar{q}})$ against mass, $m_{q\bar{q}}$, for a heavy anti-quark source.}
   \label{fig:5}
    \floatfoot{The linear nature of the graph indicates that the $q\bar{q}$ is confined for increasing $m_{q\bar{q}}$. The foot of the graph at $1$ shows that the particles remain confined even at $m_{q\bar{q}}=0$ in consistency with QCD spectrum for heavy quarks.}
\end{figure}

\begin{figure}[H]
  \centering
  \subfloat[left panel]{\includegraphics[scale=0.5]{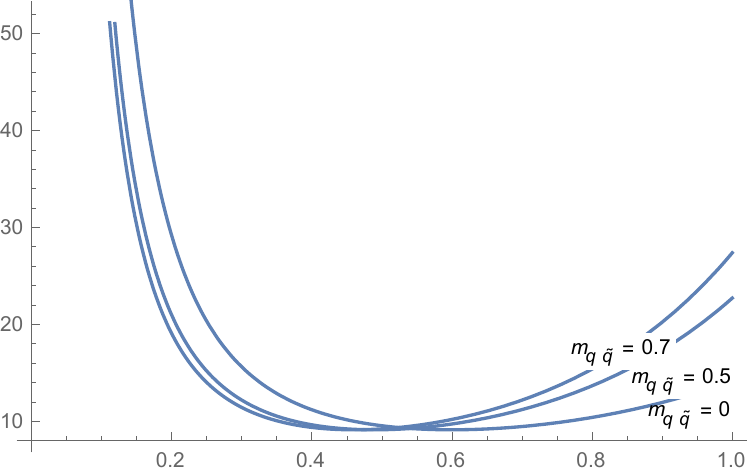}}
  \qquad
  \subfloat[right panel]{\includegraphics[scale=0.5]{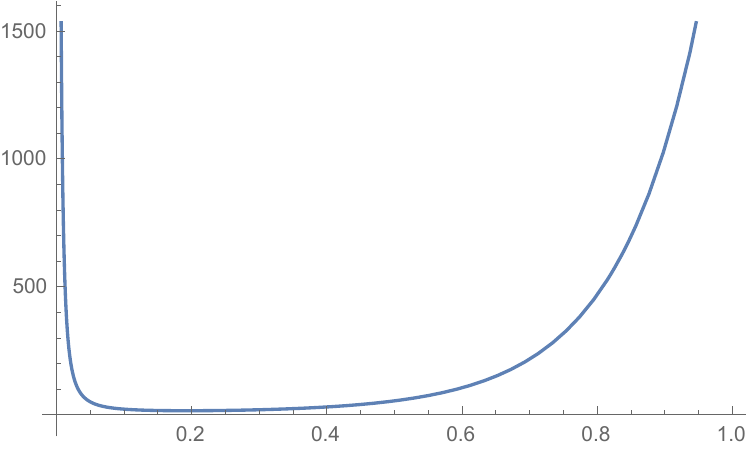}}
   \caption{A graph for color dielectric function, $G(r,m_{q\bar{q}})$, against ($r,m_{q\bar{q}}$) for specific values of $m_{q\bar{q}}$ (left) and for infinite $m_{q\bar{q}}$ (right).}
   \label{fig:6}
    \floatfoot{ {The graph in the left panel is for different values of} $m_{q\bar{q}}$, to show how the tachyons condense with increasing mass. Even though we find some tachyon condensate at $m_{q\bar{q}}=0$, we can also see that, the condensation is faster with increasing mass. The deeper the depth of the graph, the more condensed the tachyons are and the vice versa. {The graph in the {right} panel represent the tachyon condensation for infinite $m_{q\bar{q}}$ limits. It shows stronger tachyon condensation.}}
\end{figure}

\begin{figure}[H]
  \centering
   \subfloat[Left Panel]{\includegraphics[scale=0.5]{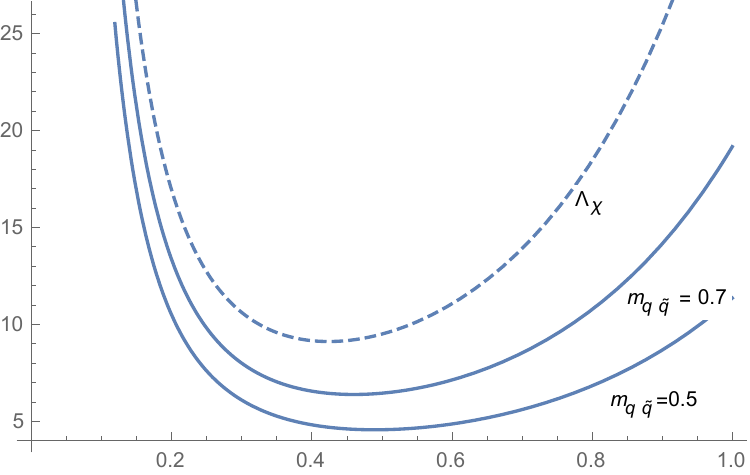}}
  \qquad
  \subfloat[Right Panel]{\includegraphics[scale=0.5]{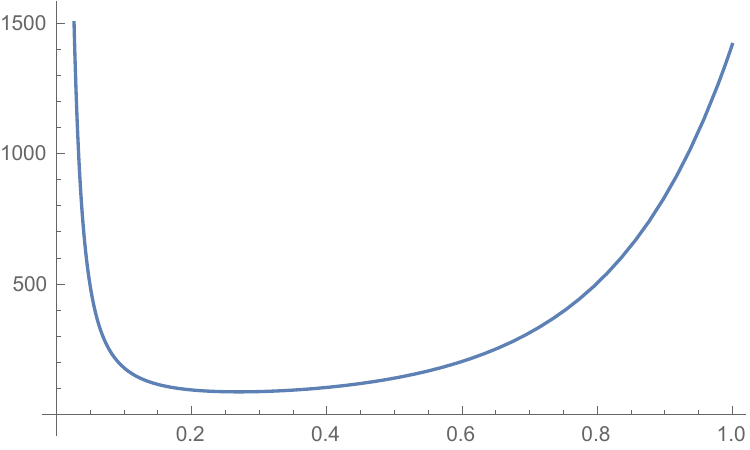}}
  
   \caption{A graph of the scalar potential $S(r,m_{q\bar{q}})$ against ($r, m_{q\bar{q}}$) for specific values of $m_{q\bar{q}}$ (left) and infinite $m_{q\bar{q}}$ (right).}
   \label{fig:7}
    \floatfoot{The potential vanishes at $m_{q\bar{q}}=0$, { for both graphs} and increases in depth as $m_{q\bar{q}}$ increases as shown in the left panel. {The graph in the right panel has a deeper depth and represents stronger confinement.} {The $S(r,m_{q\bar{q}})$ deceases with mass from the left until it attain its minimum at $S(r,m_{q\bar{q}})\rightarrow 0$ and start rising from the minimum towards the right, this is true for both graphs. The decrease implies stable confinement while the rise towards the right signifies degeneracy and screening of the static antiquark source. These are characteristics demonstrated by light mesons \cite{Gunnar}. }}
   \end{figure}

Plotting the results of equations (\ref{44}), (\ref{45}), (\ref{50b}), (\ref{scalar p}), (\ref{cornell}), (\ref{cornella}), (\ref{scalarc}) and (\ref{colorc}) in Fig.~\ref{fig:4}, Fig.~\ref{fig:5}, Fig.~\ref{fig:6}, Fig.~\ref{fig:7}, Fig.~\ref{fig:8}, Fig.~\ref{fig:9}, Fig.~\ref{fig:10} and Fig.~\ref{fig:11}, we assume that $f_\alpha=\beta=1$, $g/4\pi=\lambda=\pm1$ and $N_{c}\gg1$. Fig.~\ref{fig:4} shows graphically the relationship between the inter-quark potential $V_{c}(r,m_{q\bar{q}})$ with ($r,m_{q\bar{q}}$) for a heavy anti-quark source. The graph shows a steady increase in the gradient from $m_{q\bar{q}1}$ to $m_{q\bar{q}3}$ representing an increase in the strength of confinement from $m_{q\bar{q}1}$ to $m_{q\bar{q}3}$. Fig.~\ref{fig:5} represents the graph of $\sigma_c(m_{q\bar{q}})$ against $m_{q\bar{q}}$, it shows a linear increase in $\sigma_c(m_{q\bar{q}})$ against $m_{q\bar{q}}$ with its foot intersecting the $\sigma_c(m_{q\bar{q}})$ axis at $1$ indicating the strength of interaction even at $m_{q\bar{q}}=0$. The linearity in $\sigma_c(m_{q\bar{q}})$ depict the confinement that exist between the (anti-) quark pairs. Fig.~\ref{fig:6} shows the graph of the QCD-like vacuum which is equivalent to the tachyon potential. Tachyon condensation is related to monopole condensation which translates into confinement. The deeper the depth of the curve the more condensed the tachyons and the stronger the strength of the confinement and the vice versa. Again, Fig.~\ref{fig:7} shows the relationship between the scalar potential $S(r,m_{q\bar{q}})$ with ($r,m_{q\bar{q}}$). We notice that the greater the quark mass the deeper the depth of the curve and the smaller the minima of the curves representing faster gluon condensation. {The above discussions so far are related to the IR regime. But, same is true for Fig.~\ref{fig:8}, Fig.~\ref{fig:9}, Fig.~\ref{fig:10} and Fig.~\ref{fig:11} which incorporate some UV characteristics, the only difference is that heavy quarks ($m_{q\bar{q}}\geq 4\text{GeV}$) are required to achieve confinement in this regime due to the UV characteristics.}  
Considering all the factors discussed above, we can deduce that the vector potentials are dominant over the scalar potentials for  {light quarks in the IR regime and heavy quarks UV regime respectively.}

By way of analysis, we will compare the vector and the scalar confinement as captured in literature. The vector potential takes into account the angular momentum of the quarks whilst the scalar potentials do not. The scalar potential actually thrives on the basis that the angular momentum is partially eliminated. In the vector confinement, the energy of the quarks are carried by the gluons (string tension) whereas in the scalar confinement, the energy is centered on the quark coordinates. The scalar potential is most useful in a non rotating quark frames whilst the vector potential is a viable option for a rotating quark frames. Obviously, the `Thomas precession' that gives rise to vector confinement differs significantly from the one that gives rise to the scalar confinement \cite{massive dirac cf, schwinger, Thomas interaction, d vector-scalar}. 

   
{ \begin{figure}[H]
  \centering
   \subfloat[Left Panel]{\includegraphics[scale=0.5]{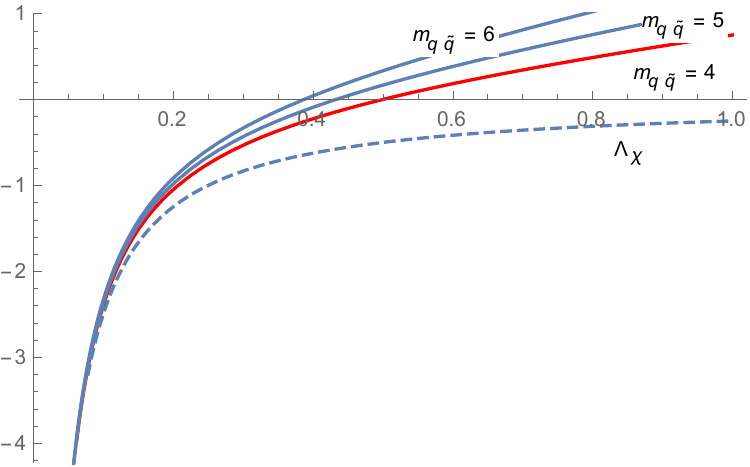}}
  \qquad
  \subfloat[Right Panel]{\includegraphics[scale=0.5]{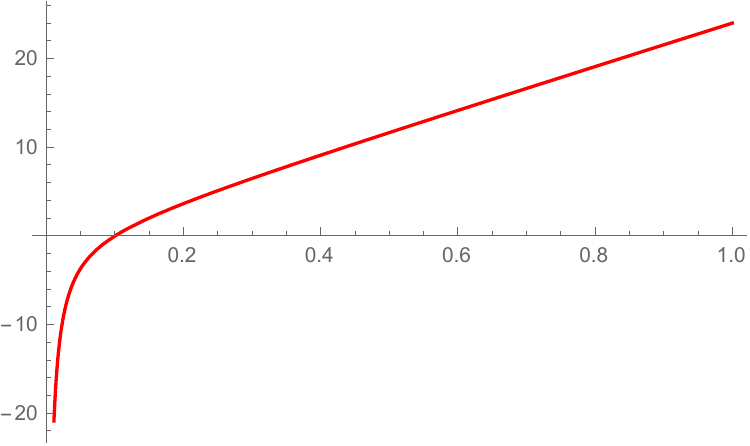}}
  
   \caption{A graph of the Cornell-like potential $V_s(r,m_{q\bar{q}})$ against ($r, m_{q\bar{q}}$) for specific values of $m_{q\bar{q}}$ (left) and infinite $m_{q\bar{q}}$ limit (right).}
   \label{fig:8}
    \floatfoot{The graph in the left panel rises steadily with increasing mass indicating strong confinement with increasing mass whilst the graph in the right panel shows stronger confinement with infinite increase in quark mass.}
   \end{figure}
   
   \begin{figure}[H]
  \centerline{\includegraphics[scale=0.5]{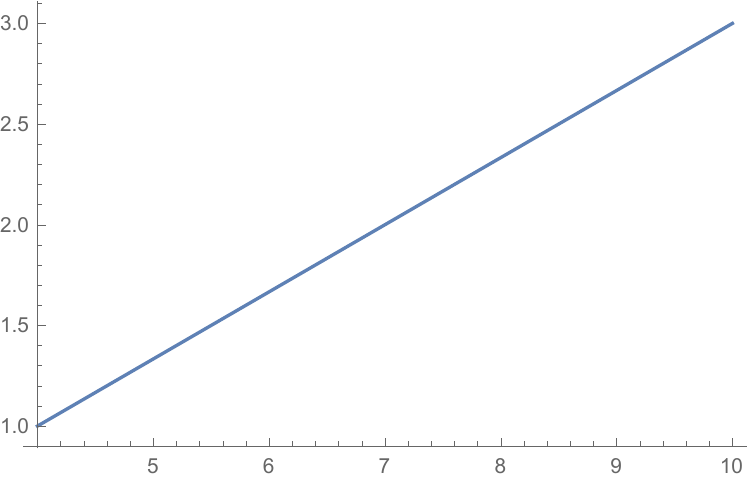}}
  \caption{A graph of string tension $\sigma_s(m_{q\bar{q}})$ resulting from the Cornell-like potential against mass, $m_{q\bar{q}}$, for a heavy quark source.}
   \label{fig:9}
    \floatfoot{The graph shows strong confinement with increasing quark mass $m_{q\bar{q}}$.}
    \end{figure}
    
    \begin{figure}[H]
  \centering
   \subfloat[Left Panel]{\includegraphics[scale=0.5]{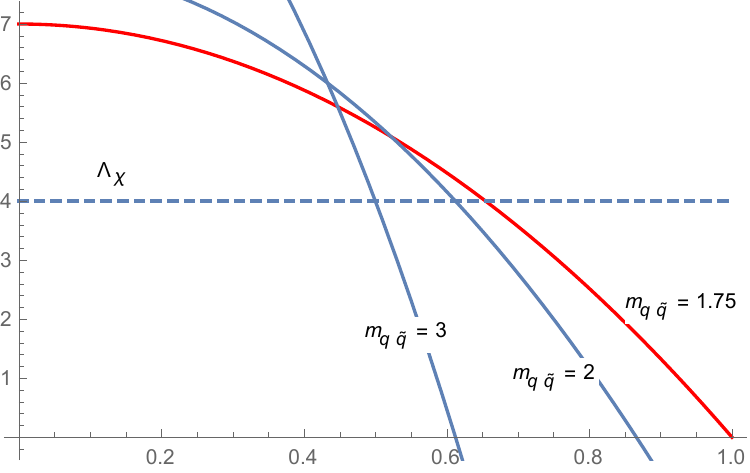}}
  \qquad
  \subfloat[Right Panel]{\includegraphics[scale=0.5]{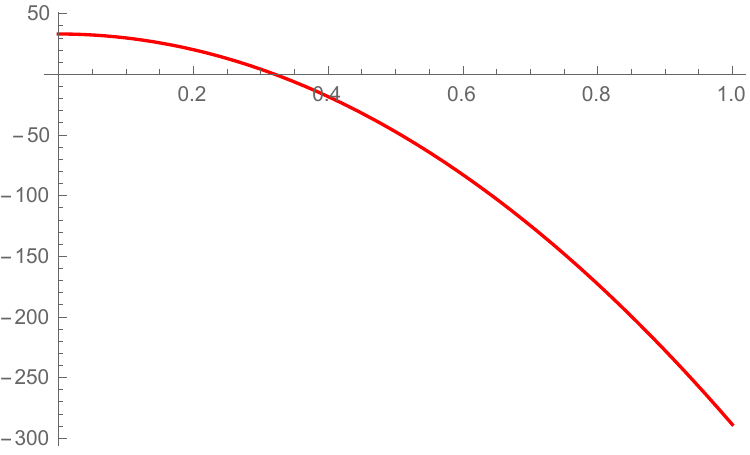}}
  
   \caption{A graph of the scalar potential $S_s(r,m_{q\bar{q}})$ against ($r, m_{q\bar{q}}$) for specific values of $m_{q\bar{q}}$ (left) and infinite $m_{q\bar{q}}$ limit (right).}
   \label{fig:10}
    \floatfoot{The potential vanishes at $m_{q\bar{q}}=0$. The scalar potential increases with increasing mass. Thus, the higher the quark mass the more likely the quarks will be confined even at higher energy regimes (small $r$) for finite quark masses as shown in the left panel. For infinite quark masses (right) the scalar potential also increases as $m_{q\bar{q}}\rightarrow\infty$, consequently confining the quarks at high energy regimes. 
    }
   \end{figure}
   \begin{figure}[H]
  \centering
   \subfloat[Left Panel]{\includegraphics[scale=0.5]{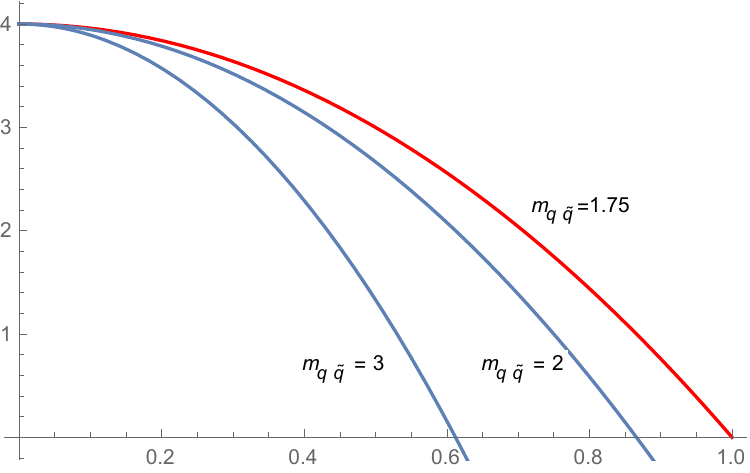}}
  \qquad
  \subfloat[Right Panel]{\includegraphics[scale=0.5]{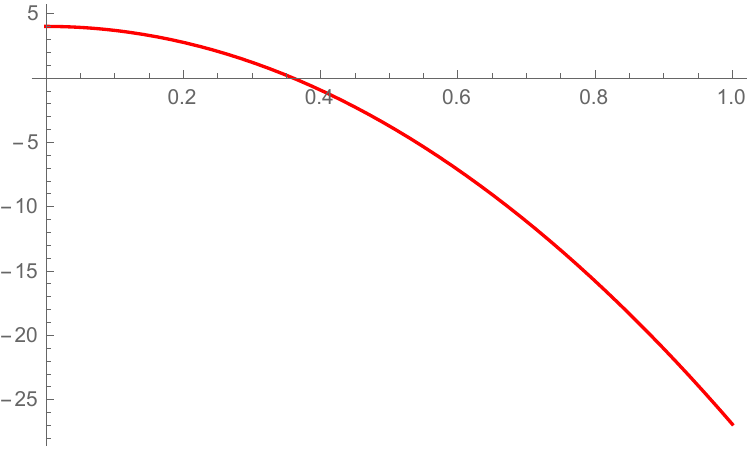}}
  
   \caption{A graph of color dielectric function $G_s(r,m_{q\bar{q}})$ against ($r, m_{q\bar{q}}$) for specific values of $m_{q\bar{q}}$ (left) and infinite $m_{q\bar{q}}$ (right).}
   \label{fig:11}
    \floatfoot{The graph in the left panel shows slight increase  in tachyon condensation with increasing quark mass, so increase in the quark mass goes into confining the quarks at higher energies $r\rightarrow 0$. Likewise, in the right panel, the tachyon condensation increases steadily as $m_{q\bar{q}}\rightarrow\infty$ resulting into confinement at higher energy regimes. 
    }
    \end{figure}
   }
\section{Conclusions}\label{conc}
The model gives an insightful details about QCD theory {in both the IR and the UV regimes} making it more efficient for consideration. We have calculated the net potential of the slowly moving quarks inside the hadron, vector potential, scalar potential energy, net potential energy and the string tensions associated with them {in both regimes}. Vector and scalar glueball masses {which are consequences of the IR regime} were also studied and their values for $f_\alpha=0.5$ (consequently, $\phi_0=500$MeV) and $\sigma_c\sim 1\text{GeV/fm}$ were calculated and compared with the existing QCD lattice results for quenched (no quark fluctuation) and unquenched (hybrid fluctuation) approximations respectively. 

The dominance of the vector potential expected in QCD theory was clearly demonstrated by computing the coupling strengths $\xi$ for both the vector and the scalar potentials. {We also find that the scalar potential (energy) must be strongly coupled in order to coexist with the vector potential (energy) which needs no coupling at all in this model framework.} 
The vector and the scalar potentials should be understood as resulting from relativistic spin-orbit corrections at short and long ranges respectively. The vector potential (energy) shows confinement of {\it chromoelectric flux} at zero mass but the scalar potential (energy) vanishes at the same mass. Thus, the vector potential (energy) is more suitable for confining light quarks whilst at least a heavy source is required to obtain scalar potential (energy). 
Furthermore, we established the relation between tachyon condensation, dual Higgs mechanism, QCD monopole condensation \cite{monopole condensation} and confinement. 
Finally, we intend to continue our series on this subject by studying confinement of fermions at a finite temperature.

\acknowledgments

We would like to thank CNPq, CAPES, and CNPq/PRONEX/FAPESQ-PB (Grant no. 165/2018) for the partial financial support. FAB acknowledges support from CNPq (Grant no. 312104/2018-9).


\end{document}